\documentclass[aps,prb,twocolumn,showpacs,amsmath,amssymb]{revtex4}
\usepackage{dcolumn}
\usepackage{bm}
\usepackage{graphicx}
\usepackage{epstopdf}
\begin{document}

\newcommand{\dpid}{$d+id'$}
\newcommand{\dmid}{$d-id'$}

\title{Chiral $d$-wave superconducting state in the core of a doubly quantized $s$-wave vortex in graphene}
\author{Annica M. Black-Schaffer}
 \affiliation{Department of Physics and Astronomy, Uppsala University, Box 516, S-751 20 Uppsala, Sweden}
 \date{\today}

\begin{abstract}
We show that the intrinsic chiral $(d_{x^2-y^2} \pm id_{xy})$-wave superconducting pairing in doped graphene is significantly strengthened in the core region of a doubly quantized $s$-wave superconducting vortex produced in a graphene-superconductor hybrid structure. The chiral $d$-wave state is induced by the proximity effect, which transfers the center-of-mass angular momentum of the $s$-wave vortex to the orbital angular momentum of the chiral $d$-wave Cooper pairs. 
The proximity effect is enhanced by the circular geometry of the vortex and we find a $[1 + (T-T_{c,J})^2]^{-1}$ temperature dependence for the chiral $d$-wave core amplitude, where $T_{c,J}$ is its intrinsic bulk transition temperature. 
We further propose to detect the chiral $d$-wave state by studying the temperature dependence of the low-energy local density of states in the vortex core, which displays a sudden radial change as function of the strength of the $d$-wave core state.
\end{abstract}
\maketitle

%
% -------------------------------------------------- %
% INTRODUCTION
% -------------------------------------------------- %
\section{Introduction}
Neutral graphene is, due to a vanishing density of states and low dimensionality, not prone to develop any symmetry breaking orders. However, several recent renormalization group studies have shown a chiral spin-singlet $d_{x^2-y^2} \pm id_{xy}$ ($d\pm id'$) state emerging from electron-electron interactions when approaching the van Hove singularity at 1/4 electron (or hole) doping.\cite{Nandkishore12, Wang11, Kiesel12} These results are consistent with investigations of strong interaction effects at lower doping levels.\cite{Black-Schaffer07, Jiang08, Honerkamp08, Pathak10, Ma11} The time-reversal symmetry breaking $d \pm id'$ combination is dictated by the sixfold symmetry of the honeycomb lattice, which makes the two $d$-wave pairing channels degenerate. \cite{Black-Schaffer07,Gonzalez08}

Experimental detection of the chiral $d$-wave state in graphene would require achieving very high doping levels, while still avoiding excessive impurity scattering, which suppresses superconductivity. A complementary approach would be to engineer a system where the intrinsic $d\pm id'$ superconducting pairing is enhanced. This would allow detection of the chiral $d$-wave state in graphene at lower doping levels and/or higher temperatures.
One obvious candidate for enhancing the intrinsic pairing is a graphene-superconductor hybrid structure, where superconductivity is induced in graphene by proximity effect. By depositing conventional spin-singlet $s$-wave superconducting contacts on graphene, proximity-induced superconductivity has already been demonstrated in graphene.\cite{Heersche07, Shailos07, Du08} Unfortunately, there is no direct coupling between an $s$-wave proximity-induced superconducting state and that of the intrinsic $d \pm id'$ pairing, due to the different orbital symmetries of the Cooper pairs.\cite{Black-Schaffer07,Black-Schaffer09} Depositing a high-temperature cuprate $d$-wave superconductor on graphene would circumvent this problem. Theoretically, $d$-wave contact Josephson junctions have been shown to significantly enhance the effect of the intrinsic $d \pm id'$-wave pairing correlations.\cite{Black-Schaffer10dSNS} However, such systems face significant material interface problems and have not yet been realized.

The fundamental problem in this case with conventional $s$-wave superconductors is the zero orbital angular momentum of the Cooper pairs. By creating a superconducting vortex with $\Delta  = |\Delta|e^{i n\theta}$, where $\theta$ is the in-plane angular coordinate, we can, however, generate a center-of-mass angular momentum $\hat{L}_z^{\rm c.m.}\Delta = n\hbar \Delta$. In the vortex core this center-of-mass angular momentum can then be transferred into an orbital angular momentum.\cite{Volovik88, Salomaa89, Sauls09} For the chiral $d \pm id'$-wave states the orbital angular momentum is $\hat{L}_z^{\rm orb}\Delta = \pm 2\hbar \Delta$. Thus, this process would require a doubly quantized $n = \pm 2$ vortex in an $s$-wave superconductor.\cite{Volovik88}

Singly quantized superconducting vortices ($n = \pm 1$) in neutral graphene have received a fair amount of attention due to the existence of two zero-energy core states,\cite{Ghaemi12} although the energy levels split when going beyond a continuum model description.\cite{Bergman09} The zero-energy core states have been argued to induce topological order in the vortex core.\cite{Ghaemi10, Herbut10} Here we instead consider $n = \pm 2$ vortices in doped graphene, and there are then, in general, no zero-energy states.\cite{Khaymovich09}
Although multiquantum vortices are less stable than singly quantized versions, they have been observed in type-I superconducting thin films at high magnetic fields.\cite{Hasegawa91} They can also be formed in superconductors having columnar or larger defects acting as pinning centers.\cite{Buzdin93, Baert95} It should therefore hopefully be experimentally feasible to generate doubly quantized vortices in a conventional $s$-wave superconductor in proximity-contact with a graphene sheet. 

In this article we establish the existence of chiral $d \pm id'$-wave superconducting states inside a doubly quantized vortex in an $s$-wave superconducting graphene sheet. Due to a strong proximity effect the $d \pm id'$-wave states survive up to temperatures of the order of the transition temperature of the proximity-induced $s$-wave order, even if the intrinsic bulk $d \pm id'$-wave transition temperature is only a very small fraction of that value. We establish the nature of this proximity effect as well as its functional temperature dependence. Furthermore, we demonstrate how the presence of the $d \pm id'$-wave core alters the low-energy local density of states in the vortex, which should be detectable by local scanning tunneling spectroscopy (STS).

%
% -------------------------------------------------- %
% METHOD
% -------------------------------------------------- %
\section{Method}
We use a nearest neighbor tight-binding Hamiltonian on the honeycomb lattice to model the $\pi$-band structure of graphene:
%
% EQUATION:
\begin{align}
\label{eq:H}
H_0 = &-t\sum_{\langle i,j\rangle,\sigma} c_{i \sigma}^\dagger c_{j \sigma} +\mu \sum_{i \sigma} c_{i \sigma}^\dagger c_{i \sigma}.
\end{align}
Here $c_{i\sigma}^\dagger$ creates an electron on the $i$th lattice site with spin $\sigma$, $t$ is the nearest neighbor hopping amplitude, and $\mu$ is the chemical potential. For simplicity we set $t = 1$ and the lattice constant $a = 1$. We further assume a high doping level in the graphene sheet, due to the external superconductor, and set $\mu = 0.5$ unless otherwise stated.

% External SC:
With the graphene sheet in close contact with a conventional spin-singlet $s$-wave superconductor, we model the proximity-induced superconducting state in graphene as
%
% EQUATION:
\begin{align}
H_U = \sum_i \Delta_U(i) c_{i \uparrow}^\dagger c_{i \downarrow}^\dagger + {\rm H.c.}.
\end{align}
We can solve the lattice Hamiltonian $H = H_0 + H_U$ within the Bogoliubov-de Gennes framework\cite{deGennesbook,Black-Schaffer08,Black-Schaffer09,Black-SchafferPRL12} by writing 
%
% EQUATION:
\begin{align}
H = X^\dagger \mathcal{H}X \ \  {\rm with} \ \ X^\dagger = (c_{i\uparrow}^\dagger,c_{i\downarrow}),
\end{align}
and then diagonalize the matrix $\mathcal{H}$ to find the eigenvalues $E^\nu$ and eigenvectors $V^\nu$, where $\nu = 1, 2, ..., 2N$ for $N$ lattice sites. By using the eigenoperators $Y^\dagger = (\gamma_\nu^\dagger)$ defined by $X = \mathcal{V}Y$, where the columns of $\mathcal{V}$ are the eigenvectors $V^\nu$, the Hamiltonian $H$ is diagonal, $H = \sum_\nu E^\nu \gamma^\dagger_\nu \gamma_\nu$, and the eigenvalues/vectors can be used to calculate any electronic property of the system. 
By assuming a constant on-site pair potential $U$ originating from the external superconductor, we can self-consistently calculate the proximity-induced pairing using the self-consistency condition $\Delta_U(i) = -U\langle c_{i\downarrow} c_{i\uparrow}\rangle$. The self-consistent loop starts with guessing $\Delta_U(i)$, diagonalizing $\mathcal{H}$ for this $\Delta_U$, using the self-consistency condition to recalculate $\Delta_U$ from the eigenvalues and eigenvectors, and then reiterating until the change in the order parameter $\Delta_U$ between two subsequent steps is less than a predetermined convergence limit.
Instead of using $U$ as a measure of the strength of the proximity-induced superconducting state we below use the experimentally more accessible transition temperature $T_{c,U}$.
To model a superconducting vortex we impose the phase rotation $\Delta_U = |\Delta_U| e^{in\theta}$ on the order parameter, with $n = \pm 2$ for a doubly quantized vortex. While computational demands limit the lattice size, and thus force a relatively short superconducting coherence length with accompanied narrow vortex cores, we have checked that our main conclusions are largely independent of the strength of the proximity-induced superconducting state. 

% Intrinsic pairing J:
%Since its repulsive origin favors a spin-singlet $d$-wave state, 
In addition to the proximity-induced superconducting state, we also assume (weak) intrinsic superconducting pairing in graphene originating from repulsive electron-electron interactions. \cite{Nandkishore12, Wang11, Kiesel12} The high density of states near the van Hove singularity efficiently screens long-range electron-electron interactions leaving only short-range repulsion, which has been shown to result in pairing that is well localized in real space. \cite{Kiesel12} With results showing no difference in the superconducting state between pairing on nearest neighbors and next nearest neighbors, even in the presence of defects such as edges,\cite{ Black-SchafferPRL12} we choose to model the intrinsic chiral $d$-wave pairing correlations as a simple spin-singlet nearest neighbor pairing:\cite{Black-Schaffer07}
 % EQUATION:
\begin{align}
\label{eq:HDelta}
H_J = \sum_{i,\delta} \Delta_{J\delta}(i)(c^\dagger_{i\uparrow}c^\dagger_{i+\delta \downarrow} - c^\dagger_{i\downarrow}c^\dagger_{i+\delta \uparrow}) + {\rm H.c.},
\end{align}
where $\delta = 1,2,3$ labels the three inequivalent nearest neighbor bonds. We solve self-consistently also for the intrinsic order parameter using $\Delta_{J\delta}(i) = -J\langle c_{i\downarrow} c_{i+\delta \uparrow} - c_{i\uparrow} c_{i+\delta \downarrow} \rangle$, where $J$ is the parametrization of the strength of the intrinsic pairing. We do not assume any relationships between the three different components of $\Delta_{J\delta}$ in each unit cell and thus there are three different symmetry solutions. The favored bulk solution of $H_0+H_J$ with transition temperature $T_{c,J}$ is $\Delta_{J,d}$; a twofold degenerate state with chiral $ d \pm i d'$ symmetry in the Brillouin zone.\cite{Black-Schaffer07} Note that the appearance of chiral $d \pm id'$-wave symmetry is dictated by the symmetry of the honeycomb lattice, where the sixfold lattice symmetry forces the two fourfold $d_{x^2-y^2}$- and $d_{xy}$-wave solutions to become degenerate. %Due to their chiral nature, these solutions have an orbital angular momentum $\hat{L}_z^{orb} = \pm 2 \hbar$ of the Cooper pairs. 
%s-wave solutions:
The third bulk solution to $H_0+H_J$ is $\Delta_{J,s}$; an extended $s$-wave solution with equal order parameters on all three nearest neighbor bonds.\cite{Black-Schaffer07,Uchoa07} In the bulk, this solution only appears subdominantly or at very strong pairing $J$. However, in the presence of an external $s$-wave superconductor $H_U$, there is a very strong proximity coupling between the two $s$-wave states,\cite{Black-Schaffer07,Black-Schaffer09} such that the extended $s$-wave state can exist up to temperatures comparable to $T_{c,U}$ even if its intrinsic transition temperature is much lower, see Fig.~\ref{fig:dwave}(c). Thus, for the full system $H_{\rm tot} = H_0+H_U+H_J$, we always have both finite $\Delta_U$ and $\Delta_{J,s}$ in the bulk. It is only in regions where the proximity-induced state $\Delta_U$ is heavily suppressed, such as in a vortex core, that the $d$-wave symmetry solutions of $H_J$ have any chance of appearing. That such $d$-wave states indeed nucleate in $n = \pm 2$ vortex cores is shown below. 
The bulk superconducting energy gap found for finite $J$ exceeds that of $J = 0$ due to the strong coupling between the two $s$-wave states. Thus, when comparing solutions with and without $d$-wave vortex core states, we cannot just compare solutions with and without $J$, we need to also adjust $U$ to $U_{\rm eff}>U$ for $J = 0$, in order to regain the same bulk superconducting energy gap for the two cases.

%
% -------------------------------------------------- %
% RESULTS:
% -------------------------------------------------- %
\section{Results}
% -------------------------------------------------- %
% vortex OPs:
% -------------------------------------------------- %
\subsection{Vortex order parameters}
\begin{figure}[thb]
\includegraphics[scale = 1.0]{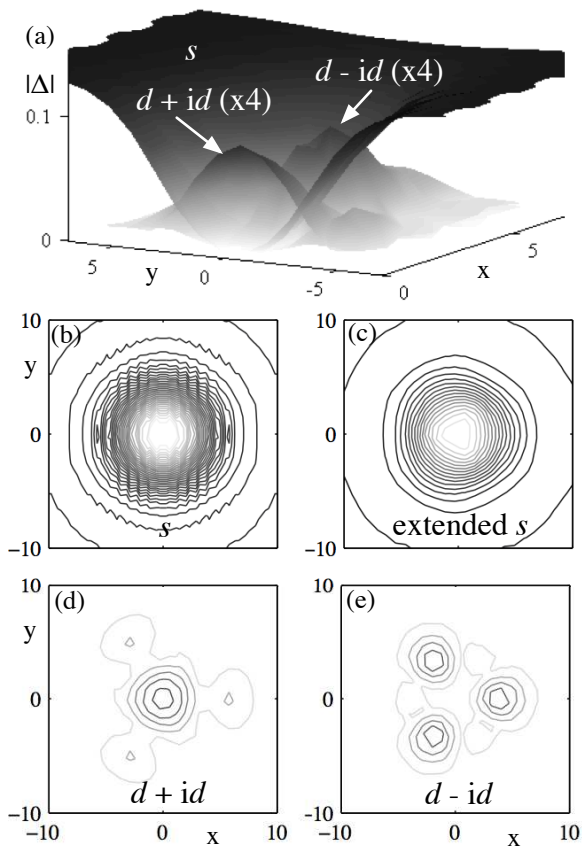}
\caption{\label{fig:amps} (a) Order parameter amplitudes in an $n = 2$ proximity-induced $s$-wave vortex in graphene. The $d$-wave amplitudes have been multiplied by $4$. Contour plots of the amplitudes of the (b) $\Delta_U$ $s$-wave, (c) $\Delta_{J,s}$ extended $s$-wave, (d) $\Delta_{J,d}$ \dpid-wave, and (e) $\Delta_{J,d}$ \dmid-wave order parameters with line spacing $0.004$ and with white lines representing zero. Phase winding around the vortex is $4\pi$ for both $s$ states, 0 for the \dpid\ state, and $8\pi$ for the \dmid\ states. Here the temperature $T = 0$, $U = 1.7$, $\mu = 0.5$, and $J = 0.88$, corresponding to $T_{c,J} = 0.1T_{c,U}$.
}
\end{figure}

We solve $H_{\rm tot} = H_0 + H_U + H_J$ self-consistently for both $\Delta_U$ and $\Delta_J$ on a finite honeycomb lattice with an imposed $4\pi$ vortex rotation for $\Delta_U$ within the Bogoliubov-de Gennes framework \cite{deGennesbook} as described above. 
% FIG 1:
This generates an $n = 2$ vortex in the $\Delta_U$ order parameter, as seen in Figs.~\ref{fig:amps}(a-b). 
%
% FIGURE:
The amplitude of the order parameter in the core region of an $n = 2$ vortex decreases as $|\Delta(r)| \sim r^2$,  in contrast to the narrower, linear decrease found for $n = 1$,\cite{Rainer96, Virtanen99} and is evident in the parabolic shape of the $s$-wave vortex profile in Fig.~\ref{fig:amps}(a). 
For $J>0$ we also find a finite $\Delta_{J,s}$, as seen in Fig.~\ref{fig:amps}(c), with the same $4\pi$ phase winding around the core as $\Delta_U$. This is due to the very strong coupling between the two $s$-wave states.
% d-waves:
Even though a finite $\Delta_{J,d}$ solution exists in doped graphene at $T = 0$, the $d$-wave solution can only appear  when the proximity-induced $s$-wave state is heavily suppressed. In the vortex core, both $s$-wave order parameters necessarily go to zero to avoid multivalueness, but instead we find finite $d$-wave order parameters in the core, as shown in Figs.~\ref{fig:amps}(d-e), as predicted.\cite{Volovik88} For the $d+id'$ solution the $\hat{L}^{orb}_z\Delta = 2\hbar\Delta$ internal orbital angular momentum can fully absorb the $\hat{L}^{c.m.}_z\Delta = 2\hbar\Delta$ center-of-mass angular momentum associated with the $s$-wave vortex phase winding. We can thus have a finite, zero-phase winding, $d+id'$ solution in the center of the core. For the accompanied $d-id'$ solution the two angular momenta instead add to a total phase winding of $8\pi$ around the vortex core. The $d-id'$ state is therefore zero in the center of the vortex, as seen in Fig.~\ref{fig:amps}(e). 
For an $n = -2$ vortex, the role of the two chiral $d \pm id'$-wave states are naturally reversed.
The spatial threefold symmetric structure of the $d \pm id'$ states follow the nearest neighbor bond directions on which the order parameters are defined. The symmetry structure is akin to the fourfold pattern found for the chiral $p \pm ip'$ states in the core of a $d$-wave vortex in fourfold symmetric high-temperature cuprate superconductors, following a similar transfer of center-of-mass angular momentum to orbital angular momentum in the vortex core.\cite{Fogelstrom11}

% -------------------------------------------------- %
% Origin of did states:
% -------------------------------------------------- %
\subsection{Origin of the $d \pm id'$ core states}
% FIG 2:
In Fig.~\ref{fig:amps} the temperature was set to absolute zero. It is then expected that $d$-wave solutions appear in the vortex core,\cite{Volovik88} also since the transition temperature $T_{c,J}$ for intrinsic pairing is finite at finite doping. What we are interested in is if the proximity-induced $s$-wave order parameter can enhance the $d$-wave state. Clearly, $\Delta_U$ helps the $\Delta_{J,s}$ state in the bulk. However, since these two states have the same orbital symmetry, they will be experimentally nearly indistinguishable, i.e.~it will be essentially impossible to tell if the bulk gap is due only to the proximity effect or if it is also enhanced by a (weak) intrinsic pair potential $J$. For the $d$-wave states the situation is very different. The $d$-wave solutions both have a very different spatial extent, only being large in the vortex core, and they have very different physical properties due to their chiral nature.

\begin{figure}[thb]
\includegraphics[scale = 1.0]{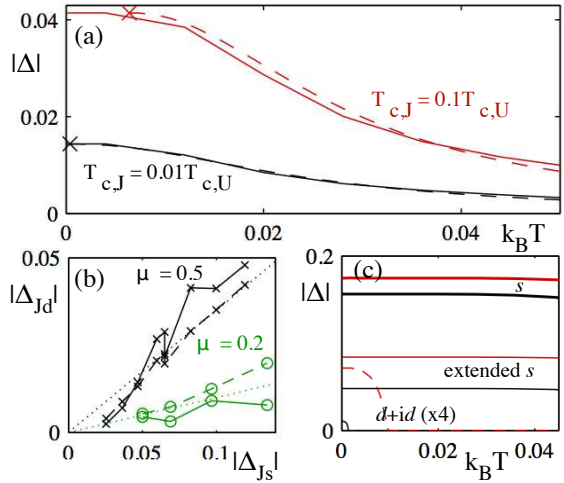}
\caption{\label{fig:dwave} (Color online) (a) Temperature dependence of the \dpid-wave core state amplitude (solid lines) for $U = 1.8$, $\mu = 0.5$, $J = 0.7$ ($T_{c,J} = 0.01T_{c,U}$) (black) and $J = 0.92$ ($T_{c,J} = 0.1T_{c,U}$) (red). Fits to Eq.~(\ref{eq:didtemp}) with $C = 1.1$ and $C = 1.4$ (dashed lines). Crosses mark the bulk $T_{c,J}$. (b) Zero-temperature amplitude of the core \dpid\ (solid lines) and \dmid\ (dashed lines) states as a function of the amplitude of the extended $s$-wave order in the bulk for $\mu = 0.5$ (black, $\times$) and $0.2$ (green, $\circ$) for multiple values of $U$ and $T_{c,J}$ in the range $(0.01 - 0.1) T_{c,U}$. Dotted lines mark linear dependence with slopes $0.1$ and 0.4, respectively. (c) Order parameters $\Delta_U$ (thick line) and $\Delta_{J,s}$ (solid) in the bulk as a function of temperature for $U = 1.8$, $\mu = 0.5$, $J = 0.7$ (black), and $J = 0.92$ (red), i.e.,~same parameters as in (a). $U = 0$ instead gives only finite $\Delta_{J,d}$ (dashed line, multiplied by 4).
}
\end{figure}

In Fig.~\ref{fig:dwave} we show the temperature dependence of the maximum amplitude of the \dpid\ core state, with crosses marking the bulk $T_{c,J} = 1\%$ and $10\%$ of $T_{c,U}$ for the two cases displayed.
% FIGURE:
 There is a remarkable enhancement of the \dpid\ (and \dmid, not shown) state far beyond its intrinsic transition temperature. Despite the, per-design weak, intrinsic pairing, there exists a finite \dpid\ state in the vortex core up to temperatures comparable to the proximity-induced transition temperature. Since the intrinsic pairing can be assumed to be weak in graphene, with a very low $T_{c,J}$, the fact that a \dpid\ state is present even far above $T_{c,J}$ might be of crucial importance for experimental detection.

% General GL for cylindrical hole:
The \dpid\ core state and its temperature dependence are due to the proximity effect in a circular hole region. Starting from the Ginzburg-Landau free energy for a superconducting order parameter $\Psi$, and studying a normal metallic (N) circular  hole region of radius $R_0$ inside a superconductor (S), we find 
 % EQUATION:
\begin{align}
\label{eq:Psieq}
\frac{1}{r}\frac{d}{dr}\left( r \frac{d \Psi(r)}{d r}\right) = \frac{1}{\xi_N^2}\Psi(r)
\end{align}
for the order parameter in the normal hole region, where $\xi_N$ is the decay length of $\Psi$ in the normal region. Equation (\ref{eq:Psieq}) has solutions $\Psi(r) \propto J_0 (i r/\xi_N)$, where $J$ is a Bessel function of the first kind.
As a boundary condition we use $\Psi(R_0) = A$, with $A$ being the value of the order parameter just inside the perimeter of the hole region. With this, we find the order parameter in the middle of the hole to be $\Psi(0) = A/J_0(iR_0/\xi_N) \approx A[1 + (R_0/\xi_N)^2]^{-1}$ for $\xi_N \gg R_0$. 
% Special for graphene:
For a vortex core in graphene we can approximate $R_0 \sim \xi_S = \hbar v_F/E_g$, with $v_F$ being the Fermi velocity and $E_g$ the superconducting energy gap.\cite{Khaymovich09} Furthermore, with the normal hole region in fact being a weak superconductor above its transition temperature $T_{c,J}$, the decay length in the clean limit is $\xi_N = \hbar v_F/[2\pi k_B (T-T_{c,J})]$.\cite{Covaci06,Black-Schaffer10dSNS} We would thus expect the temperature dependence of the $d$-wave state in the vortex core to be of the form
% EQUATION:
\begin{align}
\label{eq:didtemp}
\Delta_{J,d}(T) = \frac{\Delta_{J,d}(T = 0)}{1 + \left(\frac{2\pi C(T-T_{c,J})}{E_g}\right)^2}.
\end{align}
Here $C$ is a constant of order 1, accounting for the vortex size in units of $\xi_S$. Also, based on the numerical results in Fig.~\ref{fig:dwave}(a) we have replaced $\Delta_{J,d}(T=T_{c,J})$ with $\Delta_{J,d}(T=0)$.
In Fig.~\ref{fig:dwave}(a), the dashed lines are fits to Eq.~(\ref{eq:didtemp}) with $C = 1.1$ (black) and $C=1.4$ (red). We thus find  excellent agreement between the numerically established temperature dependence and a simple proximity-effect model. 

% Proximity effect origin:
Beyond the temperature dependence we also need to establish the origin of the finite \dpid\ state at zero temperature. In Fig.~\ref{fig:dwave}(b) we plot the maximum of the $d+id'$ (and $d-id'$) amplitude in the vortex core as a function of the amplitude of the extended $s$-wave state in the bulk far from the vortex. We find a direct, approximately linear, relationship between these two quantities, although the proportionality constant is dependent on the doping level.
We thus conclude that the finite $d+ id'$ state in the vortex core is a result of a multiple-step proximity effect. 
First, the external $s$-wave superconductor induces a uniform $s$-wave state in the graphene. Then this uniform $s$-wave state couples strongly to the intrinsic extended $s$-wave state $\Delta_{J,s}$. The strong coupling between these two states is shown by the lack of temperature dependence for $\Delta_{J,s}$, valid for essentially all temperatures below $T_{c,U}$, as seen in Fig.~\ref{fig:dwave}(c). Finally, the extended $s$-wave state is transformed by the proximity effect into the $d + id'$ state in the vortex core, with the vortex phase winding rotating into the orbital part of the Cooper pairs. 

The $d \pm id'$ states are heavily enhanced due to the above-described multiple-step proximity process, creating finite $d \pm id'$ core states far above the intrinsic $T_{c,J}$. Figure \ref{fig:dwave}(c) shows the large difference in temperature dependence between the bulk $d \pm id'$ states and both $s$-wave states, which by proximity effect determine the core $d \pm id$ states through Eq.~(\ref{eq:didtemp}). 
Large proximity effects have been observed and modeled before in so-called weak superconducting links for $T > T_c$,\cite{Golubov04,Bozovic04,Covaci06} including graphene SNS junctions with $d$-wave superconducting contacts. \cite{Black-Schaffer10dSNS}
The present case is different in two regards. First, the circular geometry of the vortex creates an even more enhanced proximity effect. In a linear SN junction there is an exponential decay of the order parameter into the N region, whereas for a circular geometry, there is a power-law decay. Second, the vortex is created in an $s$-wave superconductor, whereas the proximity-enhanced core state has a $d\pm id$-wave symmetry. The mismatch in orbital angular momentum of the Cooper pairs is compensated by different phase windings around the vortex core. 

% -------------------------------------------------- %
% Lowest QP core energies and states:
% -------------------------------------------------- %
\subsection{Quasiparticle core states}
We have so far established the presence of chiral $d$-wave core states in a doubly quantized $s$-wave vortex in graphene, even far above the intrinsic transition temperature for the $d$-wave states. Often a subdominant superconducting order parameter nucleating in the center of a vortex core gaps the low-energy vortex quasiparticles. It can thus, in principle, be detected through the lack of vortex core states, using e.g.~STS. The $d \pm id'$ states are fully gapped superconductors in the bulk but they have gapless chiral edge modes.\cite{Volovik97, Laughlin98, Senthil99} 
Thus, while the $d \pm id'$ states gap the original vortex quasiparticles, they reintroduce low-energy states in the vortex core through their edge states. We therefore do not expect any characteristic signals in the energy levels of the vortex core from a finite chiral $d$-wave core state. However, we numerically find a notable spatial difference between the original low-energy vortex core states and the $d \pm id'$ edge states. 
For an $n = 2$ $s$-wave vortex, the lowest-energy core state forms a ring structure with a finite radius $R_{\rm core}  = R_0/\sqrt{2} \sim \xi_S$, where $R_0$ is the vortex radius.\cite{Virtanen99} 
%There are core states at the center of the vortex, but those have higher energies. 
A finite $d \pm id'$ state will push this core state to higher energies, at the same time introducing low-energy chiral edge states.
Since the $d \pm id'$ states reside firmly within the vortex core, we find that the edge states have a radius $R_{\rm edge} < R_{\rm core}$ at $T = 0$. As $T$ increases, the $d \pm id'$ states become slowly weaker and the edge states start to hybridize with each other across the $d \pm id'$ domain, resulting in higher energies. At some temperature the radius of the lowest energy state will therefore jump from $R_{\rm edge}$ to $R_{\rm core}$. 

%
% FIGURE:
\begin{figure}[thb]
\includegraphics[scale = 1.0]{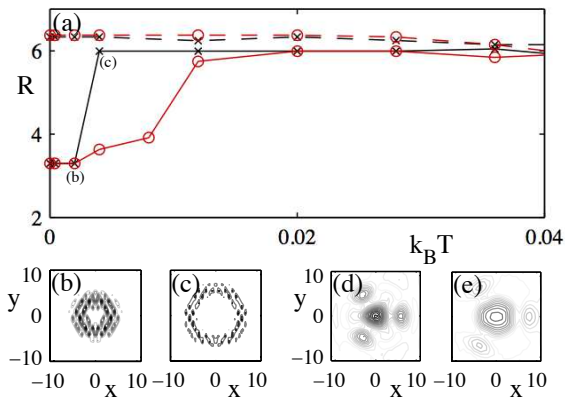}
\caption{\label{fig:lowestE} (a) Average radius of the lowest energy state in the vortex core as a function of temperature for $J = 0.67$ ($T_{c,J} = 0.01T_{c,U}$) (black, $\times$), $J = 0.7$ ($T_{c,J} = 0.015T_{c,U}$) (red, $\circ$), when $U = 1.7$, $\mu = 0.5$. Corresponding curves for $J = 0$ (dashed), using $U_{\rm eff}$ such that the bulk energy gap is unchanged. (b,c) Contour plots of the lowest energy state for $J = 0.67$ at the temperatures indicated in (a) with line spacings 0.0005 and white lines representing zero. (d,e) Contour plots of the \dpid\ amplitude at the same two temperatures as (b,c) with line spacings 0.001.
}
\end{figure}
The temperature behavior of the radius of the lowest energy state is displayed in Figs.~\ref{fig:lowestE}(a-c). The dashed lines show the radius for $J = 0$ ($R_{\rm core}$), whereas solid lines show the radius for finite $J$, where both systems have the same bulk energy gap. We clearly see a sharp jump in the radius for the finite $J$ system with increasing temperature, jumping from a radius defined by the $d \pm id'$ edge states to that of the vortex core state. This drastic change in radii happens at a temperature roughly ten times the intrinsic $T_{c,J}$, which makes for favorable experimental conditions. 
In Figs.~\ref{fig:lowestE}(d,e) the \dpid\ order parameter for temperatures just before and after the radius jump is shown. There is a notable change in the order parameter profile, but the \dpid\ state still exists for the higher temperature, even though its edge state is clearly no longer the lowest energy state. 
The relatively small radii reported in Fig.~\ref{fig:lowestE} are a consequence of the strong superconducting state. For more realistic $s$-wave vortex cores, both $R_{\rm core} \sim \xi_S$ and $R_{\rm edge} < R_{\rm core}$ will be larger. However, we still expect a sharp jump in the radius of the lowest energy state when the gapless $d \pm id'$ edge states disappear from the low-energy spectrum as the temperature is increased. This spatial change in the low-energy local density of states thus produces favorable experimental conditions for discovery of the chiral $d$-wave superconducting state in doped graphene by STS.

%
% -------------------------------------------------- %
% CONCLUSIONS:
% -------------------------------------------------- %
\section{Conclusion and discussion}
In summary we have shown that intrinsic chiral $d \pm id'$-wave superconducting pairing in doped graphene can be significantly strengthened in the core region of a doubly quantized $s$-wave superconducting vortex. 
For an $n = 2$ vortex we find a finite $d + i d'$ state in the center of the vortex core, where the center-of-mass angular momentum of the $n = 2$ phase winding has been transferred into the orbital angular momentum of the Cooper pairs of the $d + id'$ state. The $d + id'$ state is always accompanied by a $d -id'$-wave state, which is located around the center and with a phase winding of $8\pi$.

The appearance of $d \pm id'$-wave pairing in the vortex core can be described by a multiple-step proximity effect. When graphene is put in close contact with an external uniform $s$-wave superconductor, a uniform $s$-wave state is proximity-induced into the graphene. A uniform $s$-wave state couples strongly to all other possible superconducting $s$-wave states. If there are finite $d$-wave pairing correlations in a material, there is also necessarily an extended $s$-wave state possibility, which does not break fourfold rotational symmetry but in general has some $k$-space dependence. Even though the extended $s$-wave symmetry is subdominant to the $d$-wave symmetry in the bulk, it becomes favored in a graphene-superconductor hybrid structure. 
Thus the external $s$-wave superconductor also induces extended $s$-wave pairing in doped graphene. 
Even when the intrinsic transition temperature for the extended $s$-wave order is only a fraction of the transition temperature $T_{c,U}$ of the proximity-induced uniform $s$-wave order, it survives with essentially no temperature dependence up to temperatures of the order of $T_{c,U}$. In the vortex core the externally induced $s$-wave order is suppressed and with it the extended $s$-wave order. Instead, the center-of-mass angular momentum of the extended $s$-wave state is transferred into the orbital part of the Cooper pairs, producing a finite $d + id'$ state, with an accompanied $d-id'$ state. 

The strong coupling between the different $s$-wave states and the circular geometry of the vortex core give very strong $d \pm id'$ states in the core. We find that the amplitudes of the $d \pm id'$ core states have a temperature dependence $[1 +(T-T_{c,J})^2 ]^{-1}$, where $T_{c,J}$ is the intrinsic chiral $d$-wave transition temperature. There are thus significant $d \pm id'$ core amplitudes even for temperatures many times larger than $T_{c,J}$. This means that even if the $T_{c,J}$ is too low to be measurable in the bulk at experimentally achievable doping levels, its enhancement inside vortices offers an encouraging alternative route for discovery.
Doubly quantized $s$-wave superconducting vortices in doped graphene therefore present a very promising avenue for an experimental discovery of chiral $d$-wave superconductivity in graphene.
 
Since a $d \pm id'$ superconductor has gapless edge states, we find no clear signatures in the vortex energy level structure from the $d \pm id'$ core states. However, we find that there is a notable difference in the spatial distribution of the lowest-energy vortex core state when a $d \pm id'$ core is present. The lowest energy state in a doubly quantized vortex is centered around a finite radius of the order of the superconducting coherence length. In the presence of $d\pm id'$ pairing this state  is pushed up to higher energies, while instead an edge state of the chiral $d$-wave order becomes the lowest energy state. Since the $d$-wave order lives firmly within the vortex core, the edge state has a smaller radius than the original vortex core state. At some elevated temperature, up to ten times the intrinsic $d$-wave transition temperature, the chiral $d$-wave order becomes too weak to support a low-energy edge state and there is a sudden jump in the radial distribution of the lowest energy state, from the edge-state radius to that of the core-state radius. 
We therefore suggest to detect chiral $d$-wave superconductivity in doped graphene by studying the temperature dependence of the low-energy local density of states in a doubly quantized $s$-wave superconducting vortex core.

Finally, let us briefly discuss the prospects of generating an $s$-wave $n = 2$ vortex in graphene. 
In our system the $n = 2$ vortex in graphene is proximity-induced from an external superconductor. Thus, it is necessary to use an external $s$-wave superconductor which can support $n = 2$ vortices. 
Doubly quantized vortices have been observed in several different superconductors both at high magnetic fields in thin Pb films,\cite{Hasegawa91} as well as in the presence of pinning centers.\cite{Buzdin93, Baert95}
Using a magnetic field for this purpose will necessarily also induce Zeeman pair-breaking for any spin-singlet pairing, including the $d\pm id'$ core state. %, which is an effect we have not accounted for in the above calculations.
However, the critical field in type-I Pb used in Ref.~\onlinecite{Hasegawa91} is only $\sim 800$~G, thus causing a Zeeman coupling significantly smaller than 40~mK, and we therefore believe we can safely ignore this small effect in the above calculations.
Also, a $d\pm id'$ core stabilizes the $n =2$ vortex as it adds superconducting pairing to the core. This latter effect has even been suggested to make $n=2$ vortices energetically favorable over $n = 1$ vortices near the critical field and at low temperatures.\cite{Volovik88}
If instead we use pinning centers in the external superconductor to generate $n = 2$ vortices, the $d + id'$ state might be suppressed at the pinning site due to disorder scattering. We note though that the pinning site and the graphene $d + id'$ state are spatially separated, which might still make pinning centers a feasible route to generate a $d\pm id'$ vortex core state in graphene.

% -------------------------------------------------- %
% ACKNOWLEDGMENTS
% -------------------------------------------------- %
\acknowledgments
I am grateful to A.~V.~Balatsky and M.~Fogelstr\"om for discussions and the Swedish Research Council (VR) for support.

% -------------------------------------------------- %
% BIBLIOGRAPHY:
% -------------------------------------------------- %
\bibliographystyle{apsrevmy}
%\bibliography{bib2013}

\begin{thebibliography}{33}
\expandafter\ifx\csname natexlab\endcsname\relax\def\natexlab#1{#1}\fi
\expandafter\ifx\csname bibnamefont\endcsname\relax
  \def\bibnamefont#1{#1}\fi
\expandafter\ifx\csname bibfnamefont\endcsname\relax
  \def\bibfnamefont#1{#1}\fi
\expandafter\ifx\csname citenamefont\endcsname\relax
  \def\citenamefont#1{#1}\fi
\expandafter\ifx\csname url\endcsname\relax
  \def\url#1{\texttt{#1}}\fi
\expandafter\ifx\csname urlprefix\endcsname\relax\def\urlprefix{URL }\fi
\providecommand{\bibinfo}[2]{#2}
\providecommand{\eprint}[2][]{\url{#2}}

\bibitem[{\citenamefont{Nandkishore et~al.}(2012)\citenamefont{Nandkishore,
  Levitov, and Chubukov}}]{Nandkishore12}
\bibinfo{author}{\bibfnamefont{R.}~\bibnamefont{Nandkishore}},
  \bibinfo{author}{\bibfnamefont{L.~S.} \bibnamefont{Levitov}},
  \bibnamefont{and} \bibinfo{author}{\bibfnamefont{A.~V.}
  \bibnamefont{Chubukov}}, \bibinfo{journal}{Nature Phys.}
  \textbf{\bibinfo{volume}{8}}, \bibinfo{pages}{158} (\bibinfo{year}{2012}).

\bibitem[{\citenamefont{Wang et~al.}(2012)\citenamefont{Wang, Xiang, Wang,
  Wang, Yang, and Lee}}]{Wang11}
\bibinfo{author}{\bibfnamefont{W.-S.} \bibnamefont{Wang}},
  \bibinfo{author}{\bibfnamefont{Y.-Y.} \bibnamefont{Xiang}},
  \bibinfo{author}{\bibfnamefont{Q.-H.} \bibnamefont{Wang}},
  \bibinfo{author}{\bibfnamefont{F.}~\bibnamefont{Wang}},
  \bibinfo{author}{\bibfnamefont{F.}~\bibnamefont{Yang}}, \bibnamefont{and}
  \bibinfo{author}{\bibfnamefont{D.-H.} \bibnamefont{Lee}},
  \bibinfo{journal}{Phys. Rev. B} \textbf{\bibinfo{volume}{85}},
  \bibinfo{pages}{035414} (\bibinfo{year}{2012}).

\bibitem[{\citenamefont{Kiesel et~al.}(2012)\citenamefont{Kiesel, Platt, Hanke,
  Abanin, and Thomale}}]{Kiesel12}
\bibinfo{author}{\bibfnamefont{M.~L.} \bibnamefont{Kiesel}},
  \bibinfo{author}{\bibfnamefont{C.}~\bibnamefont{Platt}},
  \bibinfo{author}{\bibfnamefont{W.}~\bibnamefont{Hanke}},
  \bibinfo{author}{\bibfnamefont{D.~A.} \bibnamefont{Abanin}},
  \bibnamefont{and} \bibinfo{author}{\bibfnamefont{R.}~\bibnamefont{Thomale}},
  \bibinfo{journal}{Phys. Rev. B} \textbf{\bibinfo{volume}{86}},
  \bibinfo{pages}{020507} (\bibinfo{year}{2012}).

\bibitem[{\citenamefont{Black-Schaffer and Doniach}(2007)}]{Black-Schaffer07}
\bibinfo{author}{\bibfnamefont{A.~M.} \bibnamefont{Black-Schaffer}}
  \bibnamefont{and} \bibinfo{author}{\bibfnamefont{S.}~\bibnamefont{Doniach}},
  \bibinfo{journal}{Phys.\ Rev.\ B} \textbf{\bibinfo{volume}{75}},
  \bibinfo{pages}{134512} (\bibinfo{year}{2007}).
  
  \bibitem[{\citenamefont{Jiang}(2007)}]{Jiang08}
\bibinfo{author}{\bibfnamefont{Y.} \bibnamefont{Jiang}},
\bibinfo{author}{\bibfnamefont{D.-X.} \bibnamefont{Yao}},
\bibinfo{author}{\bibfnamefont{E. W.} \bibnamefont{Carlson}},
\bibinfo{author}{\bibfnamefont{H.-D.} \bibnamefont{Chen}},
  \bibnamefont{and} \bibinfo{author}{\bibfnamefont{J. P.}~\bibnamefont{Hu}},
  \bibinfo{journal}{Phys.\ Rev.\ B} \textbf{\bibinfo{volume}{77}},
  \bibinfo{pages}{235420} (\bibinfo{year}{2008}).
  
\bibitem[{\citenamefont{Honerkamp}(2008)}]{Honerkamp08}
\bibinfo{author}{\bibfnamefont{C.}~\bibnamefont{Honerkamp}},
  \bibinfo{journal}{Phys.\ Rev.\ Lett.} \textbf{\bibinfo{volume}{100}},
  \bibinfo{pages}{146404} (\bibinfo{year}{2008}).

\bibitem[{\citenamefont{Pathak et~al.}(2010)\citenamefont{Pathak, Shenoy, and
  Baskaran}}]{Pathak10}
\bibinfo{author}{\bibfnamefont{S.}~\bibnamefont{Pathak}},
  \bibinfo{author}{\bibfnamefont{V.~B.} \bibnamefont{Shenoy}},
  \bibnamefont{and} \bibinfo{author}{\bibfnamefont{G.}~\bibnamefont{Baskaran}},
  \bibinfo{journal}{Phys. Rev. B} \textbf{\bibinfo{volume}{81}},
  \bibinfo{pages}{085431} (\bibinfo{year}{2010}).

\bibitem[{\citenamefont{Ma et~al.}(2011)\citenamefont{Ma, Huang, Hu, and
  Lin}}]{Ma11}
\bibinfo{author}{\bibfnamefont{T.}~\bibnamefont{Ma}},
  \bibinfo{author}{\bibfnamefont{Z.}~\bibnamefont{Huang}},
  \bibinfo{author}{\bibfnamefont{F.}~\bibnamefont{Hu}}, \bibnamefont{and}
  \bibinfo{author}{\bibfnamefont{H.-Q.} \bibnamefont{Lin}},
  \bibinfo{journal}{Phys. Rev. B} \textbf{\bibinfo{volume}{84}},
  \bibinfo{pages}{121410} (\bibinfo{year}{2011}).

\bibitem[{\citenamefont{Gonz\'alez}(2008)}]{Gonzalez08}
\bibinfo{author}{\bibfnamefont{J.}~\bibnamefont{Gonz\'alez}},
  \bibinfo{journal}{Phys. Rev. B} \textbf{\bibinfo{volume}{78}},
  \bibinfo{pages}{205431} (\bibinfo{year}{2008}).

\bibitem[{\citenamefont{Heersche et~al.}(2007)\citenamefont{Heersche,
  Jarillo-Herrero, Oostinga, Vandersypen, and Morpurgo}}]{Heersche07}
\bibinfo{author}{\bibfnamefont{H.~B.} \bibnamefont{Heersche}},
  \bibinfo{author}{\bibfnamefont{P.}~\bibnamefont{Jarillo-Herrero}},
  \bibinfo{author}{\bibfnamefont{J.~B.} \bibnamefont{Oostinga}},
  \bibinfo{author}{\bibfnamefont{L.~M.~K.} \bibnamefont{Vandersypen}},
  \bibnamefont{and} \bibinfo{author}{\bibfnamefont{A.~F.}
  \bibnamefont{Morpurgo}}, \bibinfo{journal}{Nature}
  \textbf{\bibinfo{volume}{446}}, \bibinfo{pages}{56} (\bibinfo{year}{2007}).

\bibitem[{\citenamefont{Shailos et~al.}(2007)\citenamefont{Shailos, Nativel,
  Kasumov, Collet, Ferrier, Gu\'{e}ron, Deblock, and Bouchiat}}]{Shailos07}
\bibinfo{author}{\bibfnamefont{A.}~\bibnamefont{Shailos}},
  \bibinfo{author}{\bibfnamefont{W.}~\bibnamefont{Nativel}},
  \bibinfo{author}{\bibfnamefont{A.}~\bibnamefont{Kasumov}},
  \bibinfo{author}{\bibfnamefont{C.}~\bibnamefont{Collet}},
  \bibinfo{author}{\bibfnamefont{M.}~\bibnamefont{Ferrier}},
  \bibinfo{author}{\bibfnamefont{S.}~\bibnamefont{Gu\'{e}ron}},
  \bibinfo{author}{\bibfnamefont{R.}~\bibnamefont{Deblock}}, \bibnamefont{and}
  \bibinfo{author}{\bibfnamefont{H.}~\bibnamefont{Bouchiat}},
  \bibinfo{journal}{Europhys.\ Lett.} \textbf{\bibinfo{volume}{79}},
  \bibinfo{pages}{57008} (\bibinfo{year}{2007}).

\bibitem[{\citenamefont{Du et~al.}(2008)\citenamefont{Du, Skachko, and
  Andrei}}]{Du08}
\bibinfo{author}{\bibfnamefont{X.}~\bibnamefont{Du}},
  \bibinfo{author}{\bibfnamefont{I.}~\bibnamefont{Skachko}}, \bibnamefont{and}
  \bibinfo{author}{\bibfnamefont{E.~Y.} \bibnamefont{Andrei}},
  \bibinfo{journal}{Phys.\ Rev.\ B} \textbf{\bibinfo{volume}{77}},
  \bibinfo{pages}{184507} (\bibinfo{year}{2008}).

\bibitem[{\citenamefont{Black-Schaffer and Doniach}(2009)}]{Black-Schaffer09}
\bibinfo{author}{\bibfnamefont{A.~M.} \bibnamefont{Black-Schaffer}}
  \bibnamefont{and} \bibinfo{author}{\bibfnamefont{S.}~\bibnamefont{Doniach}},
  \bibinfo{journal}{Phys.\ Rev.\ B} \textbf{\bibinfo{volume}{79}},
  \bibinfo{pages}{064502} (\bibinfo{year}{2009}).

\bibitem[{\citenamefont{Black-Schaffer and
  Doniach}(2010)}]{Black-Schaffer10dSNS}
\bibinfo{author}{\bibfnamefont{A.~M.} \bibnamefont{Black-Schaffer}}
  \bibnamefont{and} \bibinfo{author}{\bibfnamefont{S.}~\bibnamefont{Doniach}},
  \bibinfo{journal}{Phys. Rev. B} \textbf{\bibinfo{volume}{81}},
  \bibinfo{pages}{014517} (\bibinfo{year}{2010}).

\bibitem[{\citenamefont{Volovik}(1988)}]{Volovik88}
\bibinfo{author}{\bibfnamefont{G.~E.} \bibnamefont{Volovik}},
 \bibinfo{journal}{J. Phys. C: Solid State Phys.} \textbf{\bibinfo{volume}{21}}, \bibinfo{pages}{L215}
  (\bibinfo{year}{1988}).
  
  \bibitem[{\citenamefont{Salomaa}(1989)}]{Salomaa89}
  \bibinfo{author}{\bibfnamefont{M.~M.} \bibnamefont{Salomaa}} \bibnamefont{and} 
\bibinfo{author}{\bibfnamefont{G.~E.} \bibnamefont{Volovik}},
 \bibinfo{journal}{J. Phys.: Condens. Matter.} \textbf{\bibinfo{volume}{1}}, \bibinfo{pages}{277}
  (\bibinfo{year}{1989}).
  
\bibitem[{\citenamefont{Sauls }(2009)}]{Sauls09}
\bibinfo{author}{\bibfnamefont{J.~A.} \bibnamefont{Sauls}} \bibnamefont{and}
  \bibinfo{author}{\bibfnamefont{M.}~\bibnamefont{Eschrig.}}, \bibinfo{journal}{New
  J. Phys.} \textbf{\bibinfo{volume}{11}}, \bibinfo{pages}{075008}
  (\bibinfo{year}{2009}).

\bibitem[{\citenamefont{Ghaemi and F.}(2012)}]{Ghaemi12}
\bibinfo{author}{\bibfnamefont{P.}~\bibnamefont{Ghaemi}} \bibnamefont{and}
  \bibinfo{author}{\bibfnamefont{F.}~\bibnamefont{Wilczek}}, \bibinfo{journal}{Phys.
  Scr.} \textbf{\bibinfo{volume}{T146}}, \bibinfo{pages}{014019}
  (\bibinfo{year}{2012}).

\bibitem[{\citenamefont{Bergman and Le~Hur}(2009)}]{Bergman09}
\bibinfo{author}{\bibfnamefont{D.~L.} \bibnamefont{Bergman}} \bibnamefont{and}
  \bibinfo{author}{\bibfnamefont{K.}~\bibnamefont{Le~Hur}},
  \bibinfo{journal}{Phys. Rev. B} \textbf{\bibinfo{volume}{79}},
  \bibinfo{pages}{184520} (\bibinfo{year}{2009}).

\bibitem[{\citenamefont{Ghaemi et~al.}(2010)\citenamefont{Ghaemi, Ryu, and
  Lee}}]{Ghaemi10}
\bibinfo{author}{\bibfnamefont{P.}~\bibnamefont{Ghaemi}},
  \bibinfo{author}{\bibfnamefont{S.}~\bibnamefont{Ryu}}, \bibnamefont{and}
  \bibinfo{author}{\bibfnamefont{D.-H.} \bibnamefont{Lee}},
  \bibinfo{journal}{Phys. Rev. B} \textbf{\bibinfo{volume}{81}},
  \bibinfo{pages}{081403} (\bibinfo{year}{2010}).

\bibitem[{\citenamefont{Herbut}(2010)}]{Herbut10}
\bibinfo{author}{\bibfnamefont{I.~F.} \bibnamefont{Herbut}},
  \bibinfo{journal}{Phys. Rev. Lett.} \textbf{\bibinfo{volume}{104}},
  \bibinfo{pages}{066404} (\bibinfo{year}{2010}).

\bibitem[{\citenamefont{Khaymovich et~al.}(2009)\citenamefont{Khaymovich,
  Kopnin, Mel'nikov, and Shereshevskii}}]{Khaymovich09}
\bibinfo{author}{\bibfnamefont{I.~M.} \bibnamefont{Khaymovich}},
  \bibinfo{author}{\bibfnamefont{N.~B.} \bibnamefont{Kopnin}},
  \bibinfo{author}{\bibfnamefont{A.~S.} \bibnamefont{Mel'nikov}},
  \bibnamefont{and} \bibinfo{author}{\bibfnamefont{I.~A.}
  \bibnamefont{Shereshevskii}}, \bibinfo{journal}{Phys. Rev. B}
  \textbf{\bibinfo{volume}{79}}, \bibinfo{pages}{224506}
  (\bibinfo{year}{2009}).

\bibitem[{\citenamefont{Hasegawa et~al.}(1991)\citenamefont{Hasegawa, Matsuda,
  Endo, Osakabe, Igarashi, Kobayashi, Naito, Tonomura, and Aoki}}]{Hasegawa91}
\bibinfo{author}{\bibfnamefont{S.}~\bibnamefont{Hasegawa}},
  \bibinfo{author}{\bibfnamefont{T.}~\bibnamefont{Matsuda}},
  \bibinfo{author}{\bibfnamefont{J.}~\bibnamefont{Endo}},
  \bibinfo{author}{\bibfnamefont{N.}~\bibnamefont{Osakabe}},
  \bibinfo{author}{\bibfnamefont{M.}~\bibnamefont{Igarashi}},
  \bibinfo{author}{\bibfnamefont{T.}~\bibnamefont{Kobayashi}},
  \bibinfo{author}{\bibfnamefont{M.}~\bibnamefont{Naito}},
  \bibinfo{author}{\bibfnamefont{A.}~\bibnamefont{Tonomura}}, \bibnamefont{and}
  \bibinfo{author}{\bibfnamefont{R.}~\bibnamefont{Aoki}},
  \bibinfo{journal}{Phys. Rev. B} \textbf{\bibinfo{volume}{43}},
  \bibinfo{pages}{7631} (\bibinfo{year}{1991}).

\bibitem[{\citenamefont{Buzdin}(1993)}]{Buzdin93}
\bibinfo{author}{\bibfnamefont{A.~I.} \bibnamefont{Buzdin}},
  \bibinfo{journal}{Phys. Rev. B} \textbf{\bibinfo{volume}{47}},
  \bibinfo{pages}{11416} (\bibinfo{year}{1993}).

\bibitem[{\citenamefont{Baert et~al.}(1995)\citenamefont{Baert, Metlushko,
  Jonckheere, Moshchalkov, and Bruynseraede}}]{Baert95}
\bibinfo{author}{\bibfnamefont{M.}~\bibnamefont{Baert}},
  \bibinfo{author}{\bibfnamefont{V.~V.} \bibnamefont{Metlushko}},
  \bibinfo{author}{\bibfnamefont{R.}~\bibnamefont{Jonckheere}},
  \bibinfo{author}{\bibfnamefont{V.~V.} \bibnamefont{Moshchalkov}},
  \bibnamefont{and}
  \bibinfo{author}{\bibfnamefont{Y.}~\bibnamefont{Bruynseraede}},
  \bibinfo{journal}{Phys. Rev. Lett.} \textbf{\bibinfo{volume}{74}},
  \bibinfo{pages}{3269} (\bibinfo{year}{1995}).

\bibitem[{\citenamefont{de~Gennes}(1966)}]{deGennesbook}
\bibinfo{author}{\bibfnamefont{P.~G.} \bibnamefont{de~Gennes}},
  \emph{\bibinfo{title}{Superconductivity of Metals and Alloys}}
  (\bibinfo{publisher}{Benjamin, New York}, \bibinfo{year}{1966}).
 
 \bibitem[{\citenamefont{Black-Schaffer and Doniach}(2008)}]{Black-Schaffer08}
\bibinfo{author}{\bibfnamefont{A.~M.} \bibnamefont{Black-Schaffer}}
  \bibnamefont{and} \bibinfo{author}{\bibfnamefont{S.}~\bibnamefont{Doniach}},
  \bibinfo{journal}{Phys.\ Rev.\ B} \textbf{\bibinfo{volume}{78}},
  \bibinfo{pages}{024504} (\bibinfo{year}{2008}).
  
  
\bibitem[{\citenamefont{Black-Schaffer}(2012)}]{Black-SchafferPRL12}
\bibinfo{author}{\bibfnamefont{A.~M.} \bibnamefont{Black-Schaffer}},
  \bibinfo{journal}{Phys. Rev. Lett.} \textbf{\bibinfo{volume}{109}},
  \bibinfo{pages}{197001} (\bibinfo{year}{2012}).
  
  \bibitem[{\citenamefont{Uchoa}(2007)}]{Uchoa07}
\bibinfo{author}{\bibfnamefont{B.}~\bibnamefont{Uchoa}} \bibnamefont{and}
  \bibinfo{author}{\bibfnamefont{A. H.}~\bibnamefont{Castro Neto}},
  \bibinfo{journal}{Phys. Rev. Lett.} \textbf{\bibinfo{volume}{98}},
  \bibinfo{pages}{146801} (\bibinfo{year}{2007}).

\bibitem[{\citenamefont{Rainer et~al.}(1996)\citenamefont{Rainer, Sauls, and
  Waxman}}]{Rainer96}
\bibinfo{author}{\bibfnamefont{D.}~\bibnamefont{Rainer}},
  \bibinfo{author}{\bibfnamefont{J.~A.} \bibnamefont{Sauls}}, \bibnamefont{and}
  \bibinfo{author}{\bibfnamefont{D.}~\bibnamefont{Waxman}},
  \bibinfo{journal}{Phys. Rev. B} \textbf{\bibinfo{volume}{54}},
  \bibinfo{pages}{10094} (\bibinfo{year}{1996}).

\bibitem[{\citenamefont{Virtanen and Salomaa}(1999)}]{Virtanen99}
\bibinfo{author}{\bibfnamefont{S.~M.~M.} \bibnamefont{Virtanen}}
  \bibnamefont{and} \bibinfo{author}{\bibfnamefont{M.~M.}
  \bibnamefont{Salomaa}}, \bibinfo{journal}{Phys. Rev. B}
  \textbf{\bibinfo{volume}{60}}, \bibinfo{pages}{14581} (\bibinfo{year}{1999}).

\bibitem[{\citenamefont{Fogelstr\"om}(2011)}]{Fogelstrom11}
\bibinfo{author}{\bibfnamefont{M.}~\bibnamefont{Fogelstr\"om}},
  \bibinfo{journal}{Phys. Rev. B} \textbf{\bibinfo{volume}{84}},
  \bibinfo{pages}{064530} (\bibinfo{year}{2011}).

\bibitem[{\citenamefont{Covaci and Marsiglio}(2006)}]{Covaci06}
\bibinfo{author}{\bibfnamefont{L.}~\bibnamefont{Covaci}} \bibnamefont{and}
  \bibinfo{author}{\bibfnamefont{F.}~\bibnamefont{Marsiglio}},
  \bibinfo{journal}{Phys.\ Rev.\ B} \textbf{\bibinfo{volume}{73}},
  \bibinfo{pages}{014503} (\bibinfo{year}{2006}).

\bibitem[{\citenamefont{Golubov et~al.}(2004)\citenamefont{Golubov, Kupriyanov,
  and Il'ichev}}]{Golubov04}
\bibinfo{author}{\bibfnamefont{A.~A.} \bibnamefont{Golubov}},
  \bibinfo{author}{\bibfnamefont{M.~Y.} \bibnamefont{Kupriyanov}},
  \bibnamefont{and} \bibinfo{author}{\bibfnamefont{E.}~\bibnamefont{Il'ichev}},
  \bibinfo{journal}{Rev. Mod. Phys.} \textbf{\bibinfo{volume}{76}},
  \bibinfo{pages}{411} (\bibinfo{year}{2004}).

\bibitem[{\citenamefont{Bozovic et~al.}(2004)\citenamefont{Bozovic, Logvenov,
  Verhoeven, Caputo, Goldobin, and Beasley}}]{Bozovic04}
\bibinfo{author}{\bibfnamefont{I.}~\bibnamefont{Bozovic}},
  \bibinfo{author}{\bibfnamefont{G.}~\bibnamefont{Logvenov}},
  \bibinfo{author}{\bibfnamefont{M.~A.~J.} \bibnamefont{Verhoeven}},
  \bibinfo{author}{\bibfnamefont{P.}~\bibnamefont{Caputo}},
  \bibinfo{author}{\bibfnamefont{E.}~\bibnamefont{Goldobin}}, \bibnamefont{and}
  \bibinfo{author}{\bibfnamefont{M.~R.} \bibnamefont{Beasley}},
  \bibinfo{journal}{Phys. Rev. Lett.} \textbf{\bibinfo{volume}{93}},
  \bibinfo{pages}{157002} (\bibinfo{year}{2004}).

\bibitem[{\citenamefont{Volovik}(1997)}]{Volovik97}
\bibinfo{author}{\bibfnamefont{G.~E.} \bibnamefont{Volovik}},
  \bibinfo{journal}{JETP Lett.} \textbf{\bibinfo{volume}{66}},
  \bibinfo{pages}{522} (\bibinfo{year}{1997}).

\bibitem[{\citenamefont{Laughlin}(1998)}]{Laughlin98}
\bibinfo{author}{\bibfnamefont{R.~B.} \bibnamefont{Laughlin}},
  \bibinfo{journal}{Phys. Rev. Lett.} \textbf{\bibinfo{volume}{80}},
  \bibinfo{pages}{5188} (\bibinfo{year}{1998}).

\bibitem[{\citenamefont{Senthil et~al.}(1999)\citenamefont{Senthil, Marston,
  and Fisher}}]{Senthil99}
\bibinfo{author}{\bibfnamefont{T.}~\bibnamefont{Senthil}},
  \bibinfo{author}{\bibfnamefont{J.~B.} \bibnamefont{Marston}},
  \bibnamefont{and} \bibinfo{author}{\bibfnamefont{M.~P.~A.}
  \bibnamefont{Fisher}}, \bibinfo{journal}{Phys. Rev. B}
  \textbf{\bibinfo{volume}{60}}, \bibinfo{pages}{4245} (\bibinfo{year}{1999}).

\end{thebibliography}

\end{document}